# Empowering Young Learners to Explore Blockchain with User-Friendly Tools: A Method Using Google Blockly and NFTs

Yun-Cheng Tsai
pecu@ntnu.edu.tw
National Taiwan Normal University
Taipei, Taiwan

Jiun-Yu Huang
yillkid@gmail.com
Small Town Intelligence Co., Ltd
Taiwan

Da-Ru Chiou
1234567si2k@gmail.com
Taitung Tung Hai Junior High School
Taiwan

## ABSTRACT

As blockchain technology continues to gain attention, there is a growing need to make it more accessible to young learners in K-12 education. However, the technical complexity and lack of accessible tools have been identified as significant barriers to adoption. Our paper proposes a new method for empowering NFTs by continuously updating their metadata using an API layer. The approach aims to reduce the barriers to entry and enable K-12 students to explore blockchain technology in the same way they learn computational thinking through visual programming tools like Scratch. Our method utilizes Google Blockly, a visual programming language, to make updating NFT metadata more accessible and engaging for young learners. By leveraging a familiar and engaging visual programming language, students can develop their computational thinking skills and explore blockchain technology in a fun and intuitive way. The paper discusses the benefits of using NFTs as a learning tool, including how they can help students understand the concept of digital ownership and value. Overall, our proposed method has the potential to promote student engagement and understanding of blockchain technology, which could have significant implications for the future of education.

## 1 INTRODUCTION

Blockchain technology and its applications have gained significant attention in recent years, and its potential to transform various industries, including education, has been widely discussed [6, 15]. However, the technical complexity and lack of accessible tools have been identified as significant barriers to its adoption in K-12 education [14]. Research has shown that the learnability of programming languages and tools is critical in promoting student engagement and understanding [4, 8].

Visual programming languages like Scratch and Google Blockly have been widely used to teach computational thinking and programming to young learners [4, 8]. Research has shown that visual programming tools can enhance students' understanding and engagement with computational thinking concepts, especially for young learners with no prior programming experience [4]. In addition, these tools have been shown to improve student learning outcomes compared to traditional text-based programming languages [1, 12].

The use of blockchain technology in education can revolutionize how we track, measure, and verify academic progress and achievement and empower students to take control of their educational records and futures. By leveraging blockchain technology, students can have a demonstrated and tamper-proof record of their educational accomplishments to share with potential employers or academic institutions. This can significantly reduce the burden on educators and institutions to maintain accurate records and enable them to spend more time teaching and researching. Additionally, blockchain-based educational platforms can help more personalized and adaptive learning experiences for students by tracking learning progress on the blockchain and adjusting teaching methods accordingly.

Research has shown that blockchain-based educational platforms can improve the efficiency and effectiveness of the educational system [3]. Furthermore, using blockchain technology in education can enhance the transparency and accuracy of academic records, reduce administrative costs, and prevent fraud [5]. Blockchain technology can also enable new models of credentialing and certification, which can benefit both students and educational institutions [10]. For example, the Open Badges project uses blockchain technology to provide verifiable digital badges for learners to showcase their skills and accomplishments [13]. Overall, the integration of blockchain technology in education has the potential to significantly improve the educational experience for students and educators alike and empower students to take control of their academic records and futures.

The use of non-fungible tokens (NFTs) is an emerging application of blockchain technology that has gained significant attention in the art world and is now being explored in other domains [2, 7]. NFTs can provide new opportunities for students to learn about digital ownership and value. However, the technical complexity and lack of accessible tools have been identified as significant barriers to the widespread adoption of blockchain technology, particularly in K-12 education [14]. Therefore, there is a need to develop user-friendly tools and approaches to enable students to explore new technologies and concepts.

Our paper proposes a new method for empowering NFTs by continuously updating their metadata using an API layer. The approach aims to reduce the barriers to entry and enable K-12 students to explore blockchain technology in the same way they learn computational thinking through visual programming tools like Scratch [8]. Our method utilizes Google Blockly, a visual programming language, to make updating NFT metadata more accessible and engaging for young learners.

The proposed method of using Google Blockly as a visual programming language to update NFT metadata aims to reduce the technical complexity of blockchain technology and make it more accessible to young learners. By leveraging a familiar and engaging visual programming language, students can develop their computational thinking skills and explore blockchain technology in a fun and intuitive way. Moreover, using visual programming languages has improved student learning outcomes compared to traditional text-based programming languages [1, 12].

Our paper discusses the benefits of using NFTs as a learning tool, including how they can help students understand the concept of digital ownership and value. It also explains the technical aspects of our proposed method and how it can be implemented in a classroom setting. Our approach could promote student engagement and understanding of blockchain technology, which could have significant implications for the future of education. In conclusion, our paper contributes to the growing literature on the potential of blockchain technology and NFTs in education. By leveraging the power of visual programming languages and innovative teaching methods, we hope to inspire a new generation of blockchain enthusiasts and foster a deeper understanding of digital ownership and value.

In the following sections of our paper, we will discuss the related work on blockchain technology and its applications in education, the details of our proposed method for using Google Blockly to update NFT metadata, the results of our implementation in a classroom setting, and a discussion of the implications and potential future directions of our work. We believe our approach can provide a new and engaging way for students to learn about blockchain technology and digital ownership while promoting their computational thinking skills. Through our research, we hope to contribute to the ongoing discussion on the potential of blockchain technology in education and inspire new approaches to teaching and learning.

## 2 RELATED WORK
## 2.1 The Impact of Blockchain Technology on Education

The core concept of value exchange in the education metaverse is achieved by implementing blockchain technology for data verification and conversion, allowing learners to obtain unique "value" and engage in exchange [11]. In this process, the education metaverse is an open and accessible learning environment that enables learners to independently generate and create value, forming a more valuable learning ecosystem through sharing and exchange.

When learners complete a learning task or achieve specific learning outcomes in the education metaverse, this data is recorded on the blockchain to create a unique token representing the learner's outstanding value. These tokens can be verified and exchanged through the blockchain. For example, learners can sell their tickets to others or use them to exchange for other learners' learning achievements or learning resources. In this process, the education metaverse provides an open exchange platform that allows learners to freely exchange, share, and create value, forming a new learning community.

This blockchain-based value exchange model in the education metaverse encourages learners to learn and generate value actively and increases learners' autonomy and self-awareness, thereby improving learning effectiveness and satisfaction. At the same time, this model also promotes innovation and knowledge sharing, forming a new knowledge community and contributing to the development and progress of education.

With the advancement of technology, education is also facing various challenges and opportunities. Among them, the education metaverse has become a new field of focus. The emergence of blockchain technology provides new possibilities for developing the education metaverse.

The core concept of blockchain technology is decentralization and distribution, which can connect various information data to form a complete data system. In the education metaverse, blockchain technology can achieve data integration and value exchange, making the education metaverse a more fair, transparent, and decentralized world.

The data integration function of blockchain technology can organically integrate and interact with data from different sources, formats, and characteristics to achieve data integrity and credibility. At the same time, the decentralization feature of blockchain technology can prevent any centralized organization or individual from controlling and dominating the development of the education metaverse, protecting user rights and creating a more fair economic environment.

In addition to data integration and decentralization, value exchange is another important application of blockchain technology. In the education metaverse, blockchain technology can achieve value exchange, allowing various information data to communicate and interact with each other, generating more value. At the same time, blockchain technology can also realize direct interaction of funds, allowing users of the education metaverse to trade on the blockchain without cumbersome centralized money flow directly. In the thinking of the education metaverse, in addition to expanding equipment, it is more important to incorporate the essential concepts of the metaverse: "creating value," "value transformation," and "concept of ownership." Through these concepts, the education metaverse can achieve the exchange of images and the transformation of thinking, enabling students to understand the value of survival when facing virtual parallel time and space. In addition, the core of the education metaverse is creating value and defining the content and form of matter humans need and are willing to exchange, thereby achieving value transformation. At the same time, the concept of ownership should also be considered, allowing students to learn how to manage, share, and exchange their importance in the education metaverse and keep up with the changes in the value of the virtual economic system under new technological transformations. The development of the education metaverse is about expanding equipment and transforming thinking, discounts, and new modes of creating value and managing ownership.

## 2.2 Incorporating Scratch into Tinker Learning for Computational Thinking

The Tinker Learning approach emphasizes hands-on, problem-solving exercises and the use of visual programming languages to engage students in active learning. Scratch, a visual programming language developed by the MIT Media Lab, has been widely used to teach computational thinking and programming to young learners [8]. Research has shown that visual programming tools like Scratch can enhance students' understanding and engagement with computational thinking concepts, especially for young learners with no prior programming experience [4]. Furthermore, this approach has improved student learning outcomes compared to traditional text-based programming languages [1, 12].

Developing 21st-century skills, including critical thinking, communication, collaboration, and creativity, is becoming increasingly important in today's rapidly evolving job market. According to a report by the World Economic Forum, these skills are considered essential for success in the Fourth Industrial Revolution [9]. Therefore, incorporating Scratch into the Tinker Learning approach, which focuses on developing these skills, is well-aligned with the needs of the modern workforce.

## 2.3 Using Polygon SDK for JavaScript to Interact with a Smart Contract

Interacting with a smart contract on the Polygon blockchain using JavaScript typically involves importing the necessary libraries, defining the contract's ABI and bytecode, creating a new instance of HDWalletProvider, and then making a new instance of Web3, passing in the provider. The contract's address is then defined, and a new contract instance is created using the ABI and address. The Polygon SDK for JavaScript provides many tools and functions for working with smart contracts on the Polygon network. An example code snippet showing how to use Polygon SDK for JavaScript to interact with a smart contract on the Polygon blockchain is as follows:

```javascript
const Web3 = require('web3');
const HDWalletProvider = require('@truffle/hdwallet-provider');
const { abi, bytecode } = require('./MyContract.json');

const provider = new HDWalletProvider({
    privateKeys: ['PRIVATE_KEY'],
    providerOrUrl: 'https://rpc-mainnet.maticvigil.com',
});

const web3 = new Web3(provider);

const contractAddress = '0xCONTRACT_ADDRESS';

const contract = new web3.eth.Contract(abi,
    contractAddress);

// Call a function on the contract
const result = await contract.methods.myFunction().call()
    ;
console.log(result);

// Send a transaction to the contract
const accounts = await web3.eth.getAccounts();
const tx = await contract.methods.myFunction().send({
    from: accounts[0],
});
console.log(tx);
```

In this example, we first import the necessary libraries, including Web3 and HDWalletProvider. We then define the contract's ABI and bytecode. We create a new instance of HDWalletProvider, passing in our private key and the Polygon Mainnet RPC endpoint URL. We then make a new model of Web3, passing in the provider. We define the contract's address and create a new contract instance using the ABI and address. We can then call functions on the

```
contract using contract.methods.myFunction().call()
```

or send transactions to the contract using the

```
contract.methods.myFunction().send()
```

passing in the from address. This is just a basic example, and the specific details of interacting with a smart contract on the Polygon blockchain will depend on the contract itself. However, teaching K-12 students how to interact with smart contracts using JavaScript can be challenging due to the technical complexity of the process. Students may struggle with understanding the concepts of wallets, private keys, and blockchain network endpoints, which are required for interacting with a smart contract. Additionally, there is a lack of user-friendly tools and methods that enable young learners to explore blockchain technology fun and intuitively.

## 2.4 Challenges in Blockchain Education and the Need for User-friendly Tools

While blockchain technology, particularly NFTs, can potentially transform various industries, including education [6], the technical complexity and lack of accessible tools have been identified as significant barriers to its adoption in K-12 education [14]. Research has suggested that user-friendly tools and approaches are crucial to making blockchain education more accessible and engaging for young learners [16].

Non-fungible tokens (NFTs) are a type of digital asset that can represent unique items, such as artwork or collectibles. They are built on blockchain technology and have been used in various industries, including education [6]. NFTs can provide new opportunities for students to learn about digital ownership and value. However, the technical complexity of blockchain technology has been identified as a significant barrier to its adoption in K-12 education [14]. User-friendly tools and approaches, such as visual programming languages and gamification, can make blockchain education more accessible and engaging for young learners [6, 16].

Therefore, there is a need to develop innovative and user-friendly tools and methods to enable students to explore new technologies and concepts like blockchain. By leveraging the power of visual programming languages and innovative teaching methods like Tinker Learning, we hope to inspire a new generation of blockchain enthusiasts and foster a deeper understanding of digital ownership and value.

To address these challenges, developing educational resources and tools that simplify interacting with smart contracts using

JavaScript is essential. This can involve leveraging visual programming languages and gamification to make the process more engaging and accessible to young learners. Additionally, the development of innovative teaching methods like Tinker Learning can help students develop 21st-century skills, including critical thinking, communication, collaboration, and creativity, which are essential for success in the modern workforce. Adopting such approaches can inspire a new generation of blockchain enthusiasts and foster a deeper understanding of digital ownership and value.

## 3  METHODS

Our paper proposes a novel method for updating the metadata of non-fungible tokens (NFTs) using an API layer to reduce barriers to entry and enable K-12 students to explore blockchain technology. Our approach leverages Google Blockly, a visual programming language, to make updating NFT metadata more accessible and engaging for young learners. By creating a custom block in Blockly using the Blockly Developer Tools, we can create an interface that generates JavaScript and XML code for the block. This interface should allow the block's JavaScript code to interact directly with the Polygon software development kit (SDK) and smart contracts. By integrating this custom block into the BlocklyDuino editor, K-12 learners can manipulate data collected from Arduino sensors and upload it to the Polygon blockchain using the custom block.

In our paper, we present a method for creating custom blocks in Google Blockly and using them to update NFT metadata. First, we create a custom block in Google Blockly that outputs the "Hello World" message. Then, we name the block and add a dummy input before visually representing the custom block and adding it to the MyBlocks category. Next, we write a code generator function that generates the code required to update NFT metadata when the custom block is used. We modify the code to update the metadata with the desired information, export the custom block toolbox and import it into BlocklyDuino. This open-source visual programming platform generates Arduino code, and we also provide all the necessary files required for running the custom block. Finally, we import the toolbox into BlocklyDuino and verify that the custom block can be used to update NFT metadata. Our proposed method could promote student engagement and understanding of blockchain technology and NFTs, which could have significant implications for the future of education. By leveraging a familiar and engaging visual programming language like Google Blockly, we hope to inspire a new generation of blockchain enthusiasts and foster a deeper understanding of digital ownership and value.

Our paper's third author is a seventh-grade student responsible for following the teaching steps we provided to interact with Blockly and smart contracts to create an IoT plant watering and growing log. He used an NFT as a digital identity for the observed plant and regularly uploaded data to the blockchain while updating the NFT's attributes. This demonstrates that our proposed method significantly lowers the barrier to entry for K-12 students to enter the blockchain world. Currently, our focus is on validating the process, but we plan to design more new course materials that create value for blockchain applications in the future.

The Google Blockly Custom Block Tutorial is from creating the "Hello World" block and using it in BlocklyDuino. This tutorial will guide us through creating our first custom block in Google Blockly and integrating it into BlocklyDuino. The "Hello World" block will output "Hello World" in BlocklyDuino. Google Blockly demo (Custom block development tool): https://blockly-demo.appspot.com/static/demos/blockfactory/index.html. Creating the First Custom Block in Google Blockly "Hello World." The approach for creating custom blocks in Blockly is as follows:

(1) Custom block in the "Blockly Developer Tools" and name the custom block "helloworld." https://blockly-demo.appspot.com/static/demos/blockfactory/index.html.
(2) Add a dummy input and use the Block Factory to design our custom block. We can set the block's type, input, output, and other properties.
(3) Generate code. We can generate the code for it using the Block Factory. Teachers can help students add the Polygon SDK in the step.
(4) Import the custom block's XML definition into our BlocklyDuino workspace. https://qiudaru.github.io/webduino/.
(5) Use the custom block in our BlocklyDuino programs by dragging it from the toolbox and connecting it to other blocks.

Following these steps, we can create custom blocks in Blockly to add new functionality to the visual programming language.

**Figure 1: Step 1. Custom block in the "Blockly Developer Tools" and name the custom block "helloworld." https://blockly-demo.appspot.com/static/demos/blockfactory/index.html**

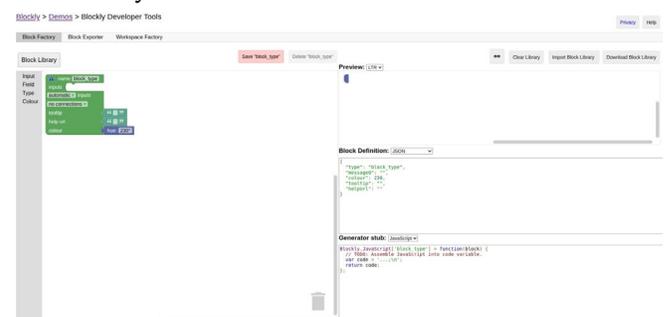

Google Blockly is a visual programming language that allows users to create JavaScript code by selecting and connecting blocks representing desired functionality [?]. This involves choosing the desired block, dragging and dropping it into the workspace, linking the blocks, configuring them by setting properties such as text, color, and tooltip, generating the JavaScript code, and exporting it to a JavaScript file or copying and pasting it into a project. Users can then test and refine the code to improve its functionality and performance. By pushing Blockly JavaScript files to a GitHub repository and configuring BlocklyDuino to access the repository's page, users can create programs that set data to a smart contract with Polygon SDK [?]. This involves creating a GitHub repository to store the Blockly JavaScript files, pushing the files to the repository,

**Figure 2: Step 2. Add a dummy input and use the Block Factory to design our custom block. We can set the block's type, input, output, and other properties.**

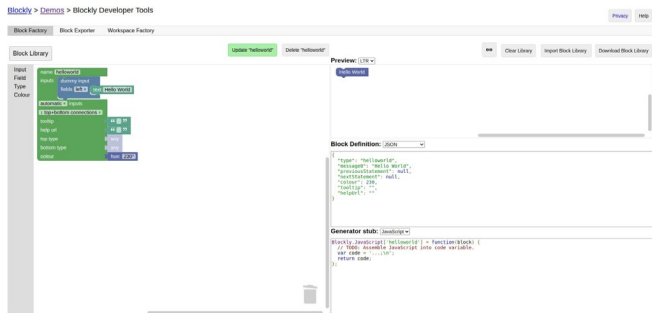

**Figure 3: Step 3. Write a post-processing tool for custom blocks and paste the "Block Definition" and "Code Generator" into it.**

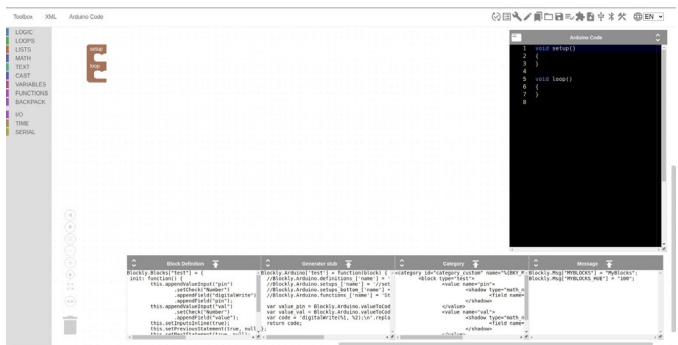

As shown in figure 2, the Block Definition on the right side defines the block.

```
Blockly.Blocks['helloworld'] = {
  init: function() {
    this.appendDummyInput()
        .appendField("Hello World");
    this.setPreviousStatement(true, null);
    this.setNextStatement(true, null);
    this.setColour(230);
 this.setTooltip("");
 this.setHelpUrl("");
  }
};
```

As shown in figure 2, the generated code in the "Generator stub" is:

```
Blockly.JavaScript['helloworld'] = function(block) {
  // TODO: Assemble JavaScript into code variable.
  var code = '...;\n';
  return code;
};
```

creating a GitHub page for the repository, configuring Blockly-Duino to access the GitHub page, creating a program in Blockly that sets data to a smart contract with Polygon SDK, generating the JavaScript code, uploading the code to an Arduino board using BlocklyDuino, connecting the board to the smart contract, and executing the program to set the data.

**Figure 4: Step 3. Add a category for custom blocks in Blockly.**

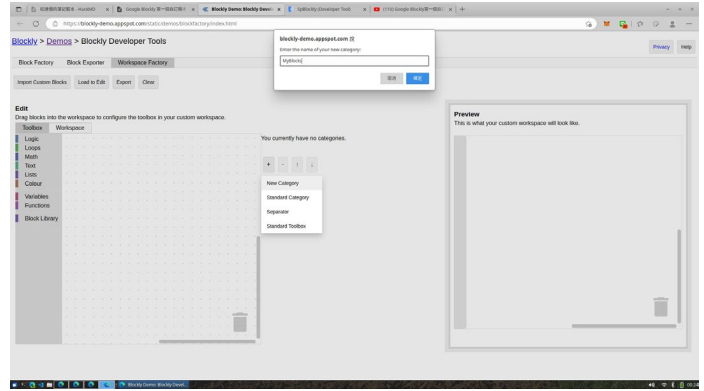

**Figure 5: Step 3. Add the custom block to the newly created category in Blockly.**

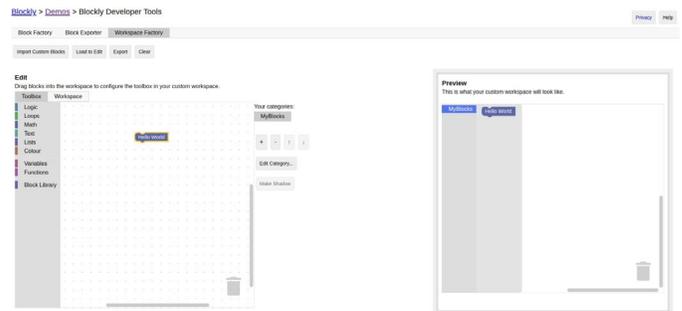

## 4   RESULTS

Our paper demonstrates the use of BlocklyDuino to control an Arduino board and obtain sensor data, with the primary challenge being to synchronize this data onto the blockchain. In response, we propose an example that uses a combination of Blockly, Blockly-Duino, Polygon SDK, and an Arduino board to create an IoT project that interacts with the blockchain. First, a smart contract is built on the Polygon blockchain, which can be accomplished using Remix or another development environment. Next, the Polygon SDK connects to the smart contract from our JavaScript code, requiring establishing a provider and a contract instance. Blockly is then employed to create a visual programming interface for our IoT device, using built-in or custom blocks to interface with sensors and other hardware. BlocklyDuino is subsequently utilized to generate the necessary Arduino code based on the blocks created in Blockly, which is then uploaded to the Arduino board. Our JavaScript code uses the Polygon SDK to send data to the smart contract whenever our IoT device collects new data, which is then stored on the blockchain for analysis or other purposes. Additionally, the Polygon

**Figure 6: Step 4. Import the XML file of the toolbox.**

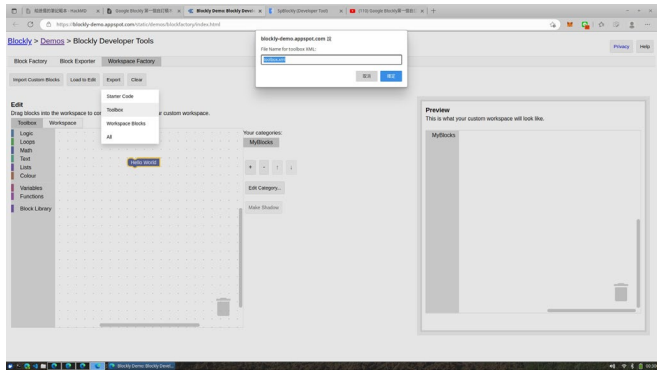

The file is as follows:

```xml
<xml xmlns="https://developers.google.com/blockly/xml" id
    ="toolbox" style="display: none">
  <category name="MyBlocks">
    <block type="helloworld"></block>
  </category>
</xml>
```

SDK can read data from the smart contract and display it in our JavaScript code or Blockly interface. Following these steps, we can develop an IoT project that interacts with the blockchain using a visual programming interface.

This is our smart contract https://goerli.etherscan.io/address/0xf4910c763ed4e47a585e2d34baa9a4b611ae448c. Our NFT is https://testnets.opensea.io/collection/untitled-collection-6860424. In conclusion, we have successfully uploaded the data collected by the sensors on the Arduino board to the Polygon testnet by following the abovementioned steps. Our experimental setup involved a plant irrigation system, where the continuously changing data related to each observed plant was updated in the metadata of an NFT that represents that plant. This project demonstrates the potential of using blockchain technology in IoT and agriculture, where tracking and managing data related to crops and farming practices is essential. Using visual programming interfaces like Blockly and BlocklyDuino, we hope to inspire more young learners to explore the possibilities of blockchain and IoT and to create innovative solutions for real-world problems. By storing irrigation records in NFTs, the sensor data, time, weather information, temperature, humidity, and crop log can be automatically updated in the attributes of the NFTs regularly. This allows students to directly experience real-time data aggregation without manual record-keeping, achieving traceability of plant growth in blockchain-based resumes. This is just a demonstration of how blockchain can be used for traceability. Blockchain technology can be used for a wide range of applications beyond resumes. By lowering the threshold for students to interact with smart contracts, they can explore the new world of blockchain economics, creating and exchanging value through innovative digital data chaining in the blockchain.

**Figure 7: Our NFT is https://testnets.opensea.io/collection/untitled-collection-6860424.**

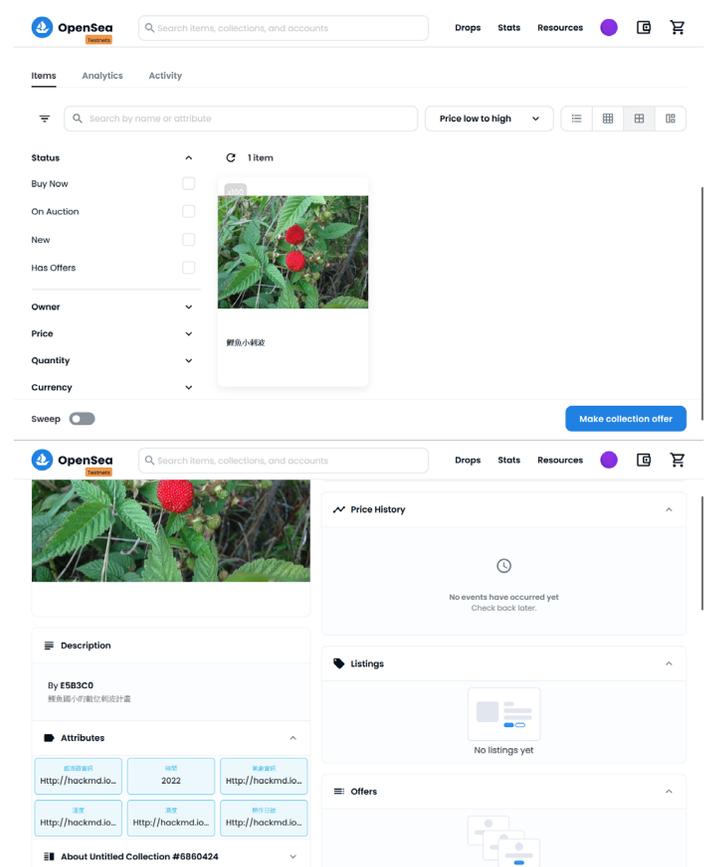

## 5 DISCUSSION

The potential of blockchain technology in the Internet of Things (IoT) context has significant implications for education. Educators can play a crucial role in helping young learners understand blockchain's advantages, particularly its ability to store data in a decentralized manner. By using a distributed network of computers to maintain a tamper-proof ledger of transactions, blockchain eliminates the need for a central authority to manage and validate data.

In the context of IoT, blockchain can store all the data generated transparently and securely, ensuring accessibility without intermediaries or gatekeepers. This allows different groups or organizations to share data, collaborate, and reduce costs. Furthermore, blockchain technology enables visual programming tools like Blockly to simplify developing and deploying IoT applications without requiring complex programming skills or hardware expertise. As blockchain technology becomes more accessible, it will create new opportunities for innovative educational programs and projects that leverage its unique features and benefits. The emergence of user-friendly block-based programming tools for teaching blockchain will empower students to explore and experiment with blockchain concepts, build decentralized applications and smart contracts, and participate in the growing blockchain ecosystem.

This will ignite a new wave of digital literacy and entrepreneurship among young learners, equipping them with the skills and knowledge to navigate and contribute to the decentralized future. One promising application of blockchain technology in education is creating a transparent and immutable system for tracking academic progress and achievement. By leveraging blockchain, students can have a verified and tamper-proof record of their educational accomplishments to share with potential employers or academic institutions. This reduces the burden on educators and institutions to maintain accurate records and enables more personalized and adaptive learning experiences.

Adopting blockchain technology in education can revolutionize how we track, measure, and verify academic progress and achievement, empowering students to take control of their educational records and futures. As educators, it is crucial to help young learners understand the implications of this technology and how it can be used to create more equitable, transparent, and collaborative ecosystems.

## 6 CONCLUSIONS

The emphasis in the education metaverse is on exchanging ideas and transforming thinking rather than simply expanding equipment. Important concepts integrated into the metaverse include "value creation," "value transformation," and "the concept of ownership," which have significant impact and meaning for personal development.

Value exchange is a core concept in the education metaverse, derived from the "effect" learners demonstrated in knowledge content, form, and meaning, which is dispersedly verified in the blockchain framework. Once the "effect" is stored as a unique token, it possesses an outstanding "value." Once learners keep value, internal motivation drives external learning behavior, generating value for others which drives personal pursuit, resulting in a repeated exchange pattern in both virtual and real worlds, like co-creation and collaboration in virtual worlds and deal in real worlds.

The value exchange has far-reaching impacts on personal development. Firstly, value exchange stimulates internal motivation, making learning an engaging, challenging, and rewarding experience rather than a forced and dull behavior. Secondly, value exchange makes it easier for individuals to recognize their value and, in turn, identify their strengths and weaknesses, developing their talents and expertise. Additionally, value exchange can help individuals build social networks, cooperate, learn, and grow with others, promoting their social skills and team spirit.

Finally, the application of value exchange in the education metaverse will be realized through blockchain technology, which allows data and funds to be linked directly through the blockchain without the need for cumbersome centralized financial flows, and how to turn this data into valuable services, making users naturally willing to pay. This not only promotes the development of education but also enables individuals to gain more value and growth. In the future education metaverse, value exchange will become an important concept, and the application of blockchain technology will become an essential tool for achieving this goal. The education metaverse will achieve human development and progress through joint efforts.

If K-12 students can reduce the threshold for entering the blockchain world and have the opportunity to participate in development and implementation through the method proposed in this article, the impact and meaning of value exchange can become even more profound. Through the realization of blockchain technology, learners and educators can exchange values more freely, co-create and extend value. This value can not only exist in the virtual world but also transform into actual wealth and social influence in real life. We look forward to the new education system bringing more possibilities and opportunities for personal development and social progress.